\documentclass[aps,twocolumn,prb,showpacs]{revtex4-1}
\usepackage{graphicx,amssymb,amsmath}
\usepackage{graphicx}
\usepackage{dcolumn}
\usepackage{bm}
\usepackage{braket}

\newcommand{\nt}{$\nu_T=1$}
\newcommand{\dl}{$d/\ell$}
\newcommand{\ec}{$e^2/\epsilon \ell$}
\newcommand{\bpar}{$B_{||}$}
\newcommand{\bperp}{$B_{\perp}$}

\newcommand{\didv}{$\partial I/\partial V$}

\begin{document}

\title{Tunneling at \nt\ in Quantum Hall Bilayers}

\author{D. Nandi$^1$, T. Khaire$^1$, A.D.K. Finck$^1$, J.P. Eisenstein$^1$, L.N. Pfeiffer$^2$, and K.W. West$^2$}

\affiliation{$^1$Condensed Matter Physics, California Institute of Technology, Pasadena, CA 91125
\\
$^2$Department of Electrical Engineering, Princeton University, Princeton, NJ 08544}

\date{\today}

\begin{abstract} Interlayer tunneling measurements in the strongly correlated bilayer quantized Hall phase at \nt\ are reported. The maximum, or critical current for tunneling at \nt, is shown to be a well-defined global property of the coherent phase, insensitive to extrinsic circuit effects and the precise configuration used to measure it, but also exhibiting a surprising scaling behavior with temperature.   Comparisons between the experimentally observed tunneling characteristics and a recent theory are favorable at high temperatures, but not at low temperatures where the tunneling closely resembles the dc Josephson effect.  The zero-bias tunneling resistance becomes extremely small at low temperatures, vastly less than that observed at zero magnetic field, but nonetheless remains finite.  The temperature dependence of this tunneling resistance is similar to that of the ordinary in-plane resistivity of the quantum Hall phase. 
\end{abstract}

\maketitle

\section{Introduction}
When the total density of electrons $n_T$ in a closely-spaced bilayer two-dimensional electron system (2DES) equals the degeneracy $eB/h$ of a single spin-resolved Landau level produced by a magnetic field $B$ perpendicular to the planes, a unique strongly correlated electron fluid emerges at low temperatures \cite{perspectives}.  Corresponding to Landau level filling fraction \nt, this quantum fluid may be equivalently described as a pseudo-spin ferromagnet or a Bose condensate of inter-layer excitons.  The essential physics underlying this unusual phase of quantum electronic matter is spontaneous inter-layer phase coherence, a broken symmetry arising from the interplay of the magnetic field and the inter- and intra-layer Coulomb interactions \cite{fertig89,wen92,yang94,moon95}.  As this phase develops, electrons in the system cease being localized in the individual layers and instead occupy a coherent linear combination of the individual layer eigenstates.  Remarkably, this delocalization occurs even in the limit of vanishingly small inter-layer tunneling.  In the clean limit, the phase $\phi$ of this linear combination in the ground state is the same for all electrons.  Spatial and temporal variations of $\phi$ cost energy and govern both the dynamics of the exciton condensate and the elementary excitations of the system. 

The coherent \nt\ bilayer exhibits several remarkable transport characteristics \cite{eisenstein13}.  When electrical currents are driven in parallel through the two layers, the bilayer behaves as a conventional quantum Hall system \cite{suen92,eisenstein92,murphy94}; the bulk of the 2DES is insulating, while a single chiral edge state at the system boundary yields a quantized Hall plateau at $\rho_{xy} = h/e^2$.  More interestingly, oppositely directed currents in the two layers are transported via the neutral exciton condensate, generating little or no dissipation or Hall voltage \cite{kellogg04,tutuc04,wiersma04}.  Recent experiments have shown that such counterflowing currents, which carry no net charge, can readily cross the insulating bulk of the 2DES \cite{finck11,nandi12,huang12}. 

Perhaps the most striking property of the coherent \nt\ bilayer state is a vast enhancement of the tunneling conductance between the two 2D layers \cite{spielman00}.  Tunneling between widely separated parallel 2D electron systems has proven to be an effective tool for the study of intra-layer electron-electron interactions both at zero and high magnetic field, $B$.  At $B=0$ the conservation of in-plane momentum leads to a resonance in the tunneling conductance \didv\ whose width is governed by the single-electron quantum lifetime \cite{murphy95,jungwirth96,zheng96}.  At high field, strong Coulomb correlations within each 2DES heavily suppress the tunneling conductance at zero bias and create broad resonances in \didv\ at voltages of order the mean Coulomb energy $e^2/\epsilon\ell$ in the system (here $\ell = (\hbar/eB)^{1/2}$ is the magnetic length) \cite{jpe92,hatsugai93,he93,johannson93,efros93}.  In both cases the observed tunneling conductance exposes the convolution of the electronic spectral functions of the two-layers \cite{finalstate}. 

At \nt\ inter-layer tunneling at low temperature undergoes a qualitative transformation when the effective layer separation \dl\ (with $d$ the physical separation between the layers) is reduced below a critical value \cite{spielman00,spielman01,eisenstein03,misra08,finck08,tiemann08a,tiemann09,yoon10,nandi12,huang12}.  The Coulombic suppression of the tunneling conductance at zero bias mentioned above is replaced by an extremely sharp peak in \didv\ which grows rapidly as \dl\ is reduced.  Eventually this peak overwhelms all other features in the tunnel spectrum.  At the lowest temperatures and \dl\ the height of this peak can exceed the tunneling conductance observed at zero magnetic field by several orders of magnitude.  In fact, the tunneling current-voltage ($IV$) characteristic at \nt\ closely resembles the dc Josephson effect observed in superconducting junctions.  It is clear that this dramatic transport anomaly is deeply reflective of the many-body physics of exciton condensation (or pseudo-ferromagnetism, if one prefers) at \nt, and there is by now an extensive theoretical literature on the subject  \cite{wen93,ezawa93,wen96,balents01,stern01,fogler01,joglekar01,iwazaki03,fertig03,bezuglyj04,abolfath04,wang04,jack04,jack05,klironomos05,wang05,bezuglyj05,fertig05,rossi05,khomeriki06,park06,fil07,eastham09,su10,sun10,eastham10,lee11,hyart11,ezawa12}.  

We here describe a series of \nt\ tunneling measurements performed in the multiply-connected Corbino geometry.  We demonstrate that while the instability and hysteresis observed in strongly tunneling devices results from extrinsic circuit effects, the maximum, or critical tunneling current, and the basic shape of the four-terminal $IV$ curve, are intrinsic properties of the coherent \nt\ phase.  We find that the critical current is remarkably insensitive to the contact configuration used to observe it and is in fact a global property of the system.  Surprisingly, the temperature dependence of the critical current, normalized by its low temperature limiting value, is found to be nearly independent of \dl.   Comparisons of the observed $IV$ curves to a recent theory \cite{hyart11} show good agreement at elevated temperatures where the $IV$ curve is smooth.  At low temperatures however, where the \nt\ tunneling most closely resembles the dc Josephson effect, the theory appears unable to capture the transition from the ``supercurrent'' branch to the resistive portion of the $IV$ curve.  Finally, we present data on the tunneling resistance along the ``supercurrent'' branch, showing it to remain finite at all temperatures.  The temperature dependence of this tunneling resistance is similar to that of the ordinary longitudinal resistivity of the \nt\ quantum Hall system, pointing to a common origin.

The plan of the paper is as follows.  Section II outlines relevant experimental details about the sample and the basic measurement techniques we have employed.   Section III presents a simple illustration of the qualitative difference between conventional intra-layer and tunneling inter-layer transport in strongly correlated bilayer 2D systems.  Section IV deals with two- versus four-terminal tunneling $IV$ characteristics at \nt, and the instabilities and hysteresis which appear when tunneling is strong.  Section V describes several of our findings concerning the maximum, or critical, tunneling current observed at \nt.  Section VI compares our tunneling data to a recent theoretical model while Sec. VII deals with the large, but finite, slope of the ``supercurrent'' branch of the $IV$ curves.  Section VIII offers a discussion of our results in the context of the general theoretical understanding of coherent \nt\ bilayer systems.  Section IX concludes the paper.
 
\section{Experimental Details}
\begin{figure}
\begin{center}
\includegraphics[width=0.8 \columnwidth] {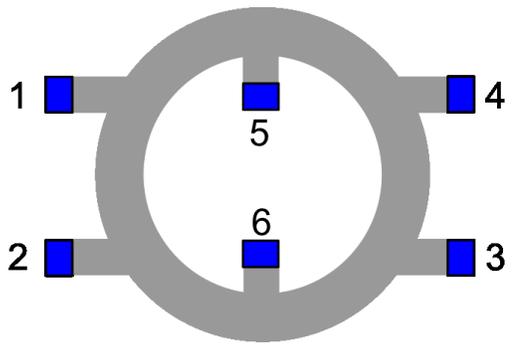}
\end{center}
\caption{(Color online) Schematic layout of a Corbino device.  The inner ring diameter is 1 mm.  Ohmic contacts, 1 - 6, are shown in blue.  Via depletion gates (not shown) these contacts can be connected to either 2D electron gas layer.  Overall top and backside gates for controlling density in the annulus are also not shown.}
\label{device}
\end{figure}

The bilayer 2D electron sample used here has been described previously \cite{finck11,nandi12}.  Grown by molecular beam epitaxy, the active region of the sample contains two 18 nm GaAs quantum wells separated by a 10 nm Al$_{0.9}$Ga$_{0.1}$As barrier.  The center-to-center quantum well separation is thus $d = 28$ nm.  As grown, each quantum well contains a 2DES of nominal density $5.5 \times 10^{10}$ cm$^{-2}$ and mobility $1 \times 10^6$ cm$^2$/Vs.  Conventional lithographic means are used to establish an annular geometry (1.4 mm outer diameter and 1.0 mm inner diameter).  Four ohmic contacts (1,2,3,4) are positioned at the ends of arms attached to the outside rim of the annulus, while two more (5,6) are located on arms attached to the inner rim.  A selective depletion scheme allows these contacts to be connected to one or the other 2D layer separately, or to remain connected to both layers simultaneously \cite{jpe90}. A schematic layout of the device is shown in Fig. \ref{device}.

Electrostatic gate electrodes on the sample top and thinned backside allow the densities $n_{1,2}$ of both 2DESs in the annulus to be separately controlled. (We here confine ourselves to the balanced case, $n_1=n_2$.)  These gates allow the effective layer separation \dl\ at \nt\ to be varied $in~situ$ at low temperatures.  The tunneling measurements reported here are primarily performed by recording the current $I$ which flows in response to a dc voltage $V_{ex}$ applied between two ohmic contacts on one or the other rim of the annulus, with one contact connected to the top 2D layer and the other to the bottom layer.  At the same time, the remaining ohmic contacts are used to record the voltage differences $V$ between the layers which develop in response to the tunneling.  Owing to the physical separation of the contacts, these voltage differences will, in general, contain both inter- and intra-layer contributions. Variations of this basic measurement protocol will be described as needed.  

To avoid confusion, we will use the subscript pair $i \alpha$ to denote the specific contacts employed in a given measurement, with $i$ referring to the contact number (1 - 6, as shown in Fig. \ref{device}) and $\alpha$ being either $t$ or $b$ to indicate which layer (top or bottom) is being contacted. For example, $V_{ex,2t,1b}$ denotes an external dc voltage applied between contact 2 on the top layer and contact 1 on the bottom layer, whereas $V_{3t,4b}$ is the measured inter-layer voltage between the top layer at contact 3 and the bottom layer at contact 4.   

\section{Interlayer versus intralayer transport}
The essential physics underlying the coherent \nt\ bilayer quantum Hall state is spontaneous interlayer phase coherence.  Owing to Coulomb interactions, electrons in the bilayer are shared equally between the layers, even in the absence of single particle tunneling.  Hence, when an electron definitely in one layer arrives at the edge of a \nt\ droplet, it is presumed to rapidly hybridize between the layers as it moves away from the injection point along the edge of the droplet.  If this hybridization occurs over a very short length scale, one might guess that the conductance measured between two contacts on the edge of the droplet would not depend much upon whether the contacts are connected to both 2D layers simultaneously, as in a conventional intra-layer transport measurement, or are connected to opposite individual layers, as in an inter-layer tunneling measurement.  As we now show, at \nt\ this guess is obviously incorrect. 

\begin{figure}
\begin{center}
\includegraphics[width=0.9 \columnwidth] {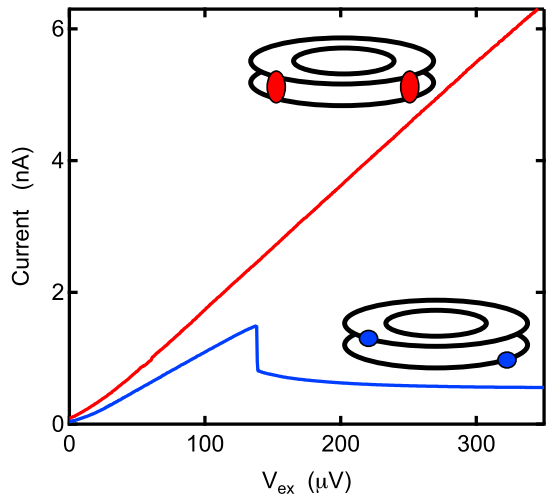}
\end{center}
\caption{(Color online) Current flowing in response to voltage $V_{ex}$ applied between contacts 1 and 2 on the outside edge of the Corbino ring at \nt.  With $d/\ell = 1.49$ and $T \sim 15$ mK, the \nt\ bilayer is deep within the coherent excitonic phase.  Red trace: Contacts connected to both layers simultaneously.  Blue trace: Contacts connected to top and bottom layers separately.}
\label{inter_vs_intra}
\end{figure}

Figure \ref{inter_vs_intra} shows two measurements at \nt\ of the current $I$ which flows in response to an external voltage $V_{ex}$ applied between contacts 1 and 2 on the outside rim of the Corbino ring.  For these measurements, the effective layer separation and temperature are $d/\ell = 1.49$ and $T \sim 15$ mK, placing the bilayer well within the coherent \nt\ phase.  For the red trace, both contacts are connected to both 2D layers simultaneously.  The data show that the current rises essentially linearly with the voltage $V_{ex}$, the slope implying a resistance of about 55 k$\Omega$. (This exceeds the value of $h/e^2=25.8$ k$\Omega$ expected for an ideal \nt\ quantum Hall droplet owing to external series resistances in the measurement circuit.) For the blue trace, where the contacts are on opposite layers, the current initially rises steadily with $V_{ex}$, albeit more slowly than in the red trace.  This is not surprising since the net series resistance (much of which is in the 2DES arms leading from the ohmic contacts in to the Corbino ring) is larger when only a single 2DES is available to carry the current toward and away from the Corbino ring.  However, at around $V_{ex} \sim 140$ $\mu$V, the current abruptly falls from about $I \sim 1.5$ nA to about $\sim 0.9$ nA.  Further increases of $V_{ex}$ elicit only a slowly falling current as the bilayer 2DES system enters a highly resistive state.  We emphasize that this highly non-linear behavior is not due to the breakdown on the quantum Hall effect itself.  Independent measurements reveal that the Hall resistance remains at its quantized value of $\rho_{xy}=h/e^2$ throughout the domain of the blue trace.

The data shown in Fig. \ref{inter_vs_intra} allow us to conclude that interlayer charge transfer, present in the blue trace but not in the red trace, fundamentally affects the two-terminal conductance along the edge of a coherent \nt\ droplet.  It seems that there is a maximum, or critical current for such tunneling transport \cite{wen93,ezawa93,spielman01,eisenstein03,tiemann08a,tiemann09,rossi05,su10,eastham10,yoon10,huang12}.  For samples in which single-particle tunneling is absent, or at least extremely weak (as it is in the present samples), it has been shown that the injection of an electron into just one of the 2D layers at the edge of the droplet necessarily excites both quasiparticle charge transport along the edge and the emission, {\it into the bulk}, of neutral excitons within the \nt\ condensate \cite{rossi05,su08,su10}.  It is this connection to the dynamics of the exciton condensate that renders inter-layer tunneling such an effective tool for studying the coherent \nt\ bilayer system.

\section{Tunneling $IV$ characteristics}
\subsection{Two- versus four-terminal tunneling measurements}
\begin{figure}
\begin{center}
\includegraphics[width=.9 \columnwidth] {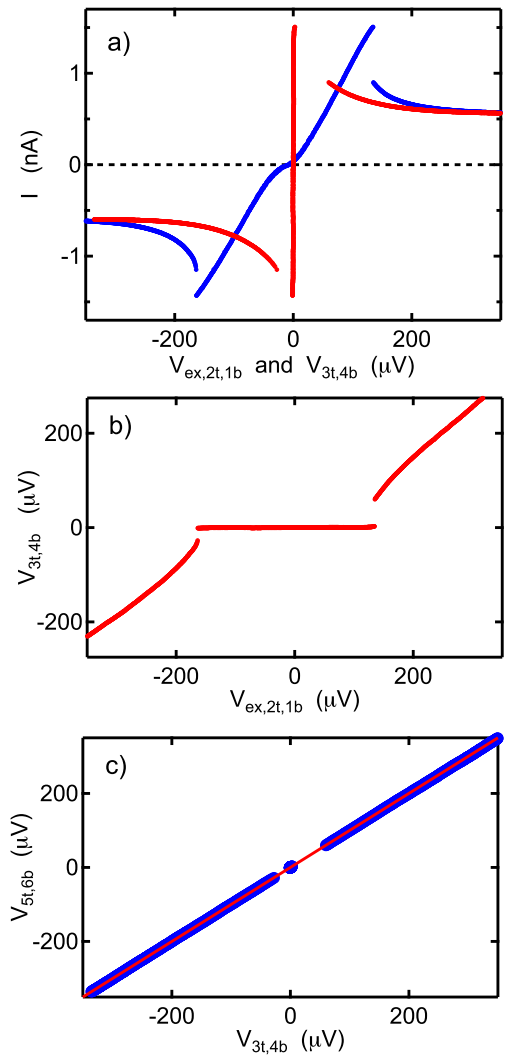}
\end{center}
\caption{(Color online) a) Two and four-terminal tunneling $IV$ curves (blue and red, respectively) at \nt\ with \dl=1.49 and $T \sim 15$ mK. b) Measured four-terminal inter-layer voltage $V_{3t,4b}$ vs. applied two-terminal voltage $V_{ex,2t,1b}$.  c) Relation of measured inter-layer voltages on the outer ($V_{3t,4b}$) and inner ($V_{5t,6b}$) rims of the annulus.  The solid red line denotes exact equality.  (Note: These data were obtained by sweeping $V_{ex,2t,1b}$ from negative to positive only.  Hence, no hysteresis is seen.)}
\label{2v4}
\end{figure}

Figure \ref{2v4}(a) contrasts two- and four-terminal tunneling $IV$ curves at \nt.  These data were again obtained at \dl\ = 1.49 and $T \sim 15$ mK where the bilayer 2DES is well within the coherent excitonic phase.  The blue trace shows the tunneling current $I$ plotted versus the dc excitation voltage $V_{ex,2t,1b}$ applied between the source and drain contacts on the outer rim of the annulus.  This is therefore a two-terminal $IV$ characteristic and, as such, includes the effects of extraneous resistances in series with the tunnel junction.  Prominent among these series resistances are ``contact'' resistances of order $h/e^2$ associated with the injection and withdrawal of current from \nt\ quantum Hall fluid in the Corbino ring itself \cite{rossi05,pesin11}, and substantial resistances in the single-layer 2D electron gas arms which connect the bilayer in the ring to the remote ohmic contacts. Taken together, the net series resistance ($R_{series} \sim 100$ k$\Omega$) accounts for the overall slope of the two-terminal $IV$ curve at low bias and its weak non-linearity \cite{finckthesis} very close to $V_{ex}=0$.

In order to remove the effects of the series resistances, a four-terminal $IV$ curve is constructed by plotting the tunneling current $I$ versus the simultaneously recorded inter-layer voltage difference $V_{3t,4b}$ which appears between the two remaining contacts on the outer rim of the Corbino ring; the result is the red curve in Fig. \ref{2v4}(a).  Plotted in this way, the tunneling $IV$ curve strongly resembles the dc Josephson effect, with a ``supercurrent'' branch on which substantial tunneling current can flow with very little, if any, voltage developing across the remote inter-layer contact pair.  Figure \ref{2v4}(b) displays the measured four-terminal voltage $V_{3t,4b}$ versus the applied voltage $V_{ex,2t,1b}$.  $V_{3t,4b}$ remains very close to zero until a critical point is reached at which it abruptly becomes finite, and the tunnel junction enters a resistive state.  Finally, Fig. \ref{2v4}(c) plots the inter-layer voltage $V_{5t,6b}$ observed on the inner rim of the Corbino device against the inter-layer voltage $V_{3t,4b}$ detected on the outer rim.  The two voltages are essentially identical when the tunnel junction is in the resistive state. The cluster of points near zero voltage arises from the ``supercurrent'' branch of the tunneling $IV$ curve; this regime requires the more careful examination presented in Sec. \ref{scbranch} below.

The Josephson-like character of tunneling at \nt\ was first noted by Spielman {\it et al.} \cite{spielman00,spielman01} using a two-terminal method.  In those earlier findings the maximum observed tunneling current was roughly 75 times smaller than that displayed in Fig. \ref{2v4}(a).  As a result, the effects of the series resistance were minimal.  In particular, the sudden jumps in the tunneling current and inter-layer voltage shown in Fig. \ref{2v4} were not observed, nor was any asymmetry in the magnitude of the positive and negative extremal tunneling current (about 5\% in Fig. \ref{2v4}a) detected.  As we next discuss, these various effects are due to circuit instabilities arising from the interplay of the series resistance and the non-linear character of tunneling at \nt.

\begin{figure}
\begin{center}
\includegraphics[width=1 \columnwidth] {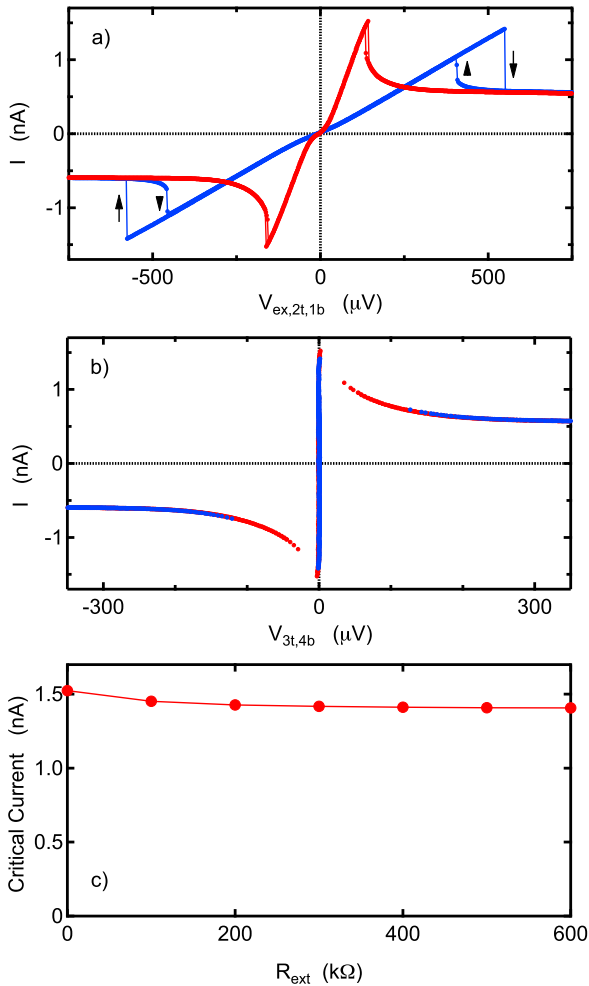}
\end{center}
\caption{(Color online) Effect of added series resistance. a) Two-terminal and b) four-terminal tunneling $IV$ curves, with (blue) and without (red) an added 300 k$\Omega$ series resistor.  c) Critical current vs. added series resistance.  All data at \nt\ with \dl\ = 1.49 and $T \sim 15$ mK.  Current source and drain on contacts 2t and 1b, respectively; four-terminal voltage probes on contacts 3t and 4b.}
\label{instab}
\end{figure} 

\subsection{Instabilities and hysteresis}
\label{instab}
Instabilities in tunneling $IV$ characteristics at \nt\ were first observed by Tiemann {\it et al.} \cite{tiemann08a} and subsequently discussed theoretically by Hyart and Rosenow \cite{hyart11}.  As observed previously \cite{spielman00,spielman01} and as is evident in Fig. \ref{2v4}, the tunneling current $I$ at \nt\ drops rapidly when the system leaves the ``supercurrent'' state and becomes highly resistive.  As the resistive state initially develops, the differential resistance $\partial V/\partial I$ of the tunnel junction is therefore negative.  If $\partial V/\partial I$ is negative and sufficiently small, or if the series resistance $R_{series}$ is sufficiently large, the two can cancel one another.   Put another way, the circuit load line $I = (V_{ex}-V)/R_{series}$ may intersect the intrinsic $I(V)$ curve of the tunnel junction at more than one point.  The circuit is then unstable and often hysteretic.  This is a common and sometimes useful aspect of circuits containing strongly non-linear elements (e.g. tunnel diode oscillators).

The basic correctness of the above explanation for the observed instabilities is easily verified by systematically adding external series resistances $R_{ext}$ to the tunneling circuit.  Figure \ref{instab}(a) contrasts the two-terminal tunneling $IV$ curves with and without an added $R_{ext}=300$ k$\Omega$ series resistor.  With the external resistor in place the slope of the two-terminal $IV$ is appropriately reduced and the jump between the ``supercurrent'' and resistive states is moved to a larger absolute $V_{ex}$.  As the figure indicates, the jump is hysteretic with the sweep direction of $V_{ex}$, with the jump from the ``supercurrent'' state to the resistive one always occurring at a larger absolute $V_{ex}$ than the jump in the reverse direction.  (As a corollary to this, when jumping from the resistive to the ``supercurrent'' branch, the junction arrives at a lower absolute current than the maximum current attained immediately prior to jumping in the reverse direction.)  While the hysteresis loop is clearly magnified by the added $R_{ext}$, it remains observable even at $R_{ext}=0$ due to the relatively large on-chip series resistance which cannot be removed from the tunneling circuit.

Owing to the circuit instability, there are gaps in the observable four-terminal $IV$ characteristic at small, but non-zero, $V$.  These gaps are widest when the tunneling is strong and/or the net series resistance is large.  Nevertheless, as Fig. \ref{instab}b shows, the four-terminal tunneling $IV$ curves obtained with and without the added external series resistor $R_{ext}$ are essentially identical wherever both are measurable.  This near-perfect agreement was verified in numerous measurements with $R_{ext}$ ranging from 0 to 600 k$\Omega$.  We therefore conclude that there is an intrinsic tunneling $IV$ characteristic at \nt, although portions of it may be unobservable due to circuit instabilities.  In particular, there appears to be a well-defined maximum, or critical, tunneling current that the coherent \nt\ bilayer system can support before significant inter-layer voltage appears between the two layers.  Figure \ref{instab}(c) shows that this maximum current is nearly independent \cite{rext} of the added external series resistance $R_{ext}$, falling by only about 8\% as $R_{ext}$ is increased from 0 to 600 k$\Omega$.

\begin{figure}
\begin{center}
\includegraphics[width=1 \columnwidth] {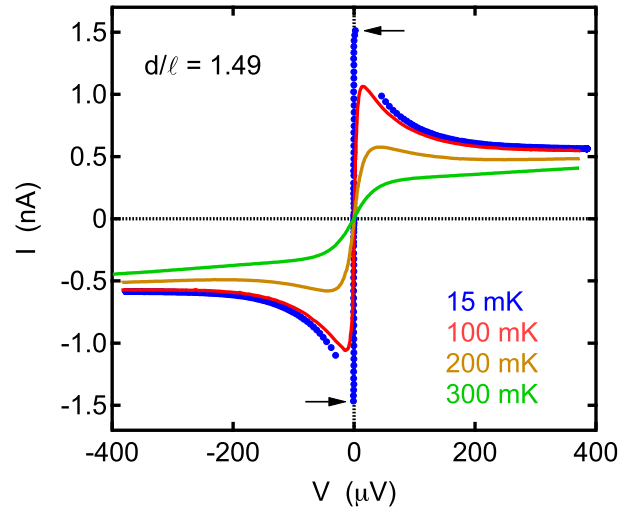}
\end{center}
\caption{(Color online) Four-terminal tunneling $IV$ curves at \nt\ and \dl=1.49 at $T = 15$, 100, 200, and 300 mK.  Note that only the $T \sim 15$ mK $IV$ curve exhibits a data gap due to the circuit instability discussed in the text.  Arrows indicate critical tunneling current at $T \sim 15$ mK.}
\label{sampleIVs}
\end{figure}

\section{Tunneling critical current}

\subsection{Empirical definition}
Figure \ref{sampleIVs} shows representative four-terminal $IV$ curves at \nt\ and effective layer separation \dl=1.49.  Traces at four temperatures, $T =15$, 100, 200, and 300 mK are shown.  For the $T = 15$ and 100 mK data there is no difficulty in identifying the critical current $I_c$.  At $T = 200$ mK the tunneling current still shows a well-defined local maximum at low voltage which can be used to define $I_c$.  However, by $T = 300$ mK no low voltage local maximum is seen in the tunneling current, even if it is visually clear that the current exhibits a smeared step discontinuity centered at $V = 0$.  Unless otherwise explicitly noted, we will assign a critical current only when the $IV$ characteristic exhibits a well-defined local maximum near $V=0$.  In those cases where the circuit instability is present, we take the critical current to be that observed just before the system jumps from the ``supercurrent'' branch to the resistive branch.

We note in passing that of all of the $IV$ curves shown in Fig. \ref{sampleIVs}, only the $T=15$ mK data display the circuit instability (and concomitant gap in the data) discussed in the previous section.  Among these $d/\ell = 1.49$ tunneling $IV$ curves, only at this lowest temperature is the tunneling strong enough that the cancellation of the series resistance by the intrinsic negative differential resistance of the junction occurs.  As discussed in Sec. \ref{HR}, continuous, non-hysteretic four-terminal tunneling $IV$ curves are observed down to the lowest temperatures in this sample, provided $d/\ell \gtrsim 1.6$.

\begin{figure}
\begin{center}
\includegraphics[width=1 \columnwidth] {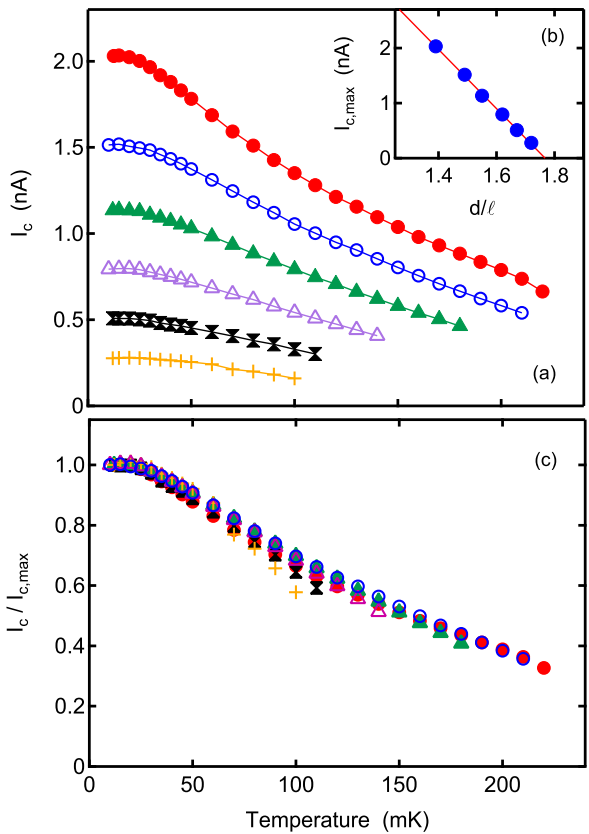}
\end{center}
\caption{(Color online) (a) \nt\ tunneling critical current $I_c$ vs. temperature. Top to bottom: \dl\ = 1.39, 1.49, 1.55, 1.62, 1.67, and 1.72.  (b) Low temperature maximum critical current $I_{c,max}$ vs. \dl.  (c) Normalized critical current $I_c/I_{c,max}$ vs. temperature for same \dl\ values as in (a). }
\label{Ic_T_dl}
\end{figure}

\subsection{Temperature and \dl\ dependence}
 Figure \ref{Ic_T_dl} summarizes the temperature and \dl\ dependence of the tunneling critical current $I_c$ at \nt.  Figure \ref{Ic_T_dl}(a) shows the temperature dependence of $I_c$ for six values of the effective layer separation: \dl\ = 1.39, 1.49, 1.55, 1.62, 1.67, and 1.72.  In each case, $I_c$ appears to saturate below about $T = 25$ mK; we use this saturation value to define $I_{c,max}$, the maximum critical current.  The inset, Fig. 6(b), reveals that $I_{c,max}$ falls essentially linearly with increasing \dl, extrapolating to zero at about $d/\ell = 1.77$.  For $d/\ell \gtrsim 1.8$ all vestiges of the coherent \nt\ quantum Hall phase are lost and the system increasingly resembles two independent 2D layers.

In Fig. \ref{Ic_T_dl}(c) the normalized critical current (defined as $I_c/I_{c,max}$) is plotted versus temperature for all six values of \dl.  Remarkably, the temperature dependence of this normalized critical current appears to be virtually independent of \dl.  Small deviations from this simple scaling behavior are, however, visible, especially at the highest \dl\ values (\dl=1.67 and 1.72). 
 
 \subsection{Contact independence}
\begin{figure}
\begin{center}
\includegraphics[width=1 \columnwidth] {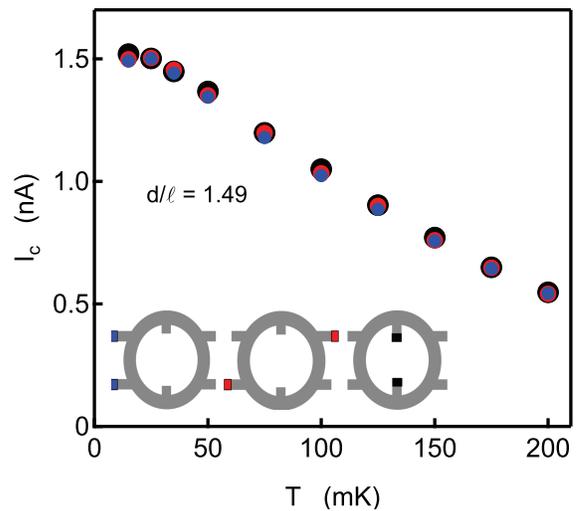}
\end{center}
\caption{(Color online) \nt\ tunneling critical current $I_c$ vs. temperature at \dl\ = 1.49 for three different source and drain contact pairs of widely different separations.  Inset: Colored squares qualitatively indicate the locations of the source and drain contacts for the three configurations.}
\label{contact_indep}
\end{figure}

Figure \ref{contact_indep} shows the \nt\ critical current $I_c$ versus temperature at \dl\ = 1.5 measured with three different source-drain contact pairs on the Corbino annulus.  The distance between the source and drain, measured along the relevant rim of the Corbino annulus, varies by more than a factor of 3 for these three contact sets.  In spite of this, the measured critical currents are virtually identical at all temperatures.  As noted previously \cite{huang12}, this observation supports the existing evidence \cite{finck08,tiemann09}  that tunneling at \nt\ occurs throughout the bulk of the 2D electron system. 

\begin{figure}
\begin{center}
\includegraphics[width=.9 \columnwidth] {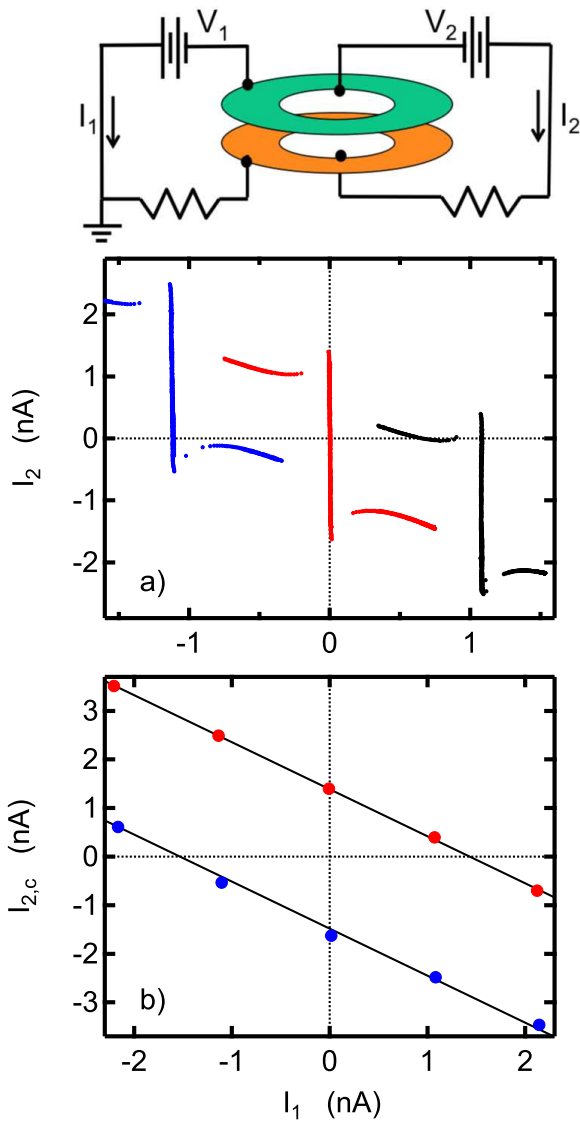}
\end{center}
\caption{(Color online) Tunneling at \nt\ with dual source-drain contacts.  The diagram illustrates the set-up.  (a) Tunneling currents $I_2$ vs. $I_1$ observed while bias voltage $V_2$ is swept and $V_1$ is held fixed. Left to right: $V_1 = -300$, 0, and +300 $\mu$V. (b) Critical values of $I_2$, indicated by arrows, vs. the stationary $I_1$ value on the ``supercurrent'' branch.  The upper diagonal line is a linear least-squares fit to the upper $I_2$ critical current; the slope is -0.97.  The lower diagonal line is parallel to the upper diagonal, but shifted down by 3.02 nA.  Data taken at $d/\ell = 1.49$ and $T \sim 15$ mK.}
\label{dualsourcedrain}
\end{figure}

\subsection{Multiple source-drain pairs}
In their experiments, Huang {\it et al} \cite{huang12} observed that when two inter-layer source-drain contact pairs, one on each rim of a Corbino annulus, are used to simultaneously inject currents $I_1$ and $I_2$ into one of the 2D layers (and withdraw them from the other), the individual currents may exceed the critical current $I_c$, provided that the sum $|I_1+I_2|$ does not. We have reproduced this result, using the set-up shown in Fig. \ref{dualsourcedrain}.

Independent inter-layer dc bias voltages $V_1$ and $V_2$ are applied between the top and bottom layer contacts on the outer and inner rim of the Corbino ring, using contact pairs (2t,1b) and (5t,6b), respectively.  In contrast to Huang {\it et al.} \cite{huang12}, only one of these bias circuits is referenced to the ground potential; the other circuit floats.  This prevents net current from flowing across the bulk of the 2DES.  The measurement consists of recording the currents $I_1$ and $I_2$ which develop as one of the bias voltages ($V_2$ for the data in Fig. \ref{dualsourcedrain}) is slowly swept while the other ($V_1$) is held fixed.  Figure \ref{dualsourcedrain}(a) plots, left to right, the three $I_2$ vs. $I_1$ characteristics observed as $V_2$ is swept at $V_1 = -300$, 0, and +300 $\mu$V.  In each case, the ``supercurrent'' branch is readily identified as the portion of the data in which $I_1$ remains constant while $I_2$ ranges between two critical values.  These two extremal $I_2$ values are plotted versus the stationary $I_1$ value in Fig. \ref{dualsourcedrain}(b).  The upper diagonal line is a linear least-squares fit to the upper critical $I_2$ value (the fitted slope is -0.97).  The lower diagonal line is parallel to the upper line, only shifted downward by 3.02 nA.  In agreement with Huang {\it et al.} \cite{huang12}, these data demonstrate that the actual critical current is determined by the sum $|I_1+I_2|$, not the individual currents.  This result vividly demonstrates that tunneling is not confined to small regions near the source and drain contacts but is instead taking place throughout the sample.

\section{Tunneling Lineshape Analysis} \label{HR}
Recently, Hyart and Rosenow (hereafter referred to as HR) have presented a theory of tunneling in disordered coherent \nt\ bilayers which allows for a quantitative comparison with experiment \cite{hyart11}.  Expanding upon earlier work by Stern {\it et al.} \cite{stern01}, HR assume that the condensate phase $\phi$ is heavily disordered, containing much quenched vorticity and possessing both a finite correlation length $\xi$ and a correlation time $\tau_\phi$.
These measures of disorder, combined with the intrinsic parameters of the clean coherent \nt\ state (the pseudo-spin stiffness $\rho_s$, the pseudo-spin-wave velocity $u$, and single particle tunnel splitting $\Delta_{SAS}$), allow HR to effectively explain the small magnitude of the observed tunneling critical currents, their scaling with sample area, the strong suppression of tunneling by an added in-plane magnetic field, and the surprisingly weak signatures of the observed \cite{spielman01} linearly-dispersing pseudo-spin-wave modes \cite{balents01,stern01,fogler01}.  Moreover, HR are led to a specific prediction for the shape of the tunneling $IV$ characteristic:   

\begin{multline}\label{eq1}
I(V) =  I_0 \int \frac{dq}{\pi(1+q^2)^{3/2}}\Big[ \frac{\alpha}{\alpha^2+(V/V_0-q)^2}\\
-\frac{\alpha}{\alpha^2+(V/V_0+q)^2}\Big].
\end{multline}

Hence, in the HR theory three parameters, $I_0$, $V_0$, and $\alpha$, determine the shape of the $IV$ curve. Roughly speaking, $I_0$ and $V_0$ determine the maximum tunneling current \cite{equation} and the voltage beyond which the current becomes small. The parameter $\alpha = \xi/u\tau_\phi$ encodes the impact of the correlation time. HR argue that most of the temperature dependence of the $IV$ curve arises from $\alpha$.  In the limit $\alpha \ll 1$, which suffices for our purposes, Eq. (1) reduces to:

\begin{equation}
I(V)=\frac{2I_0}{\pi(1+(V/V_0)^2)^{3/2}}{\rm arctan}(V/\alpha V_0).
\end{equation}
The product $\alpha V_0$ thus sets the voltage scale for the initial rise of the current toward its maximum value.  

\begin{figure}
\begin{center}
\includegraphics[width=.9 \columnwidth] {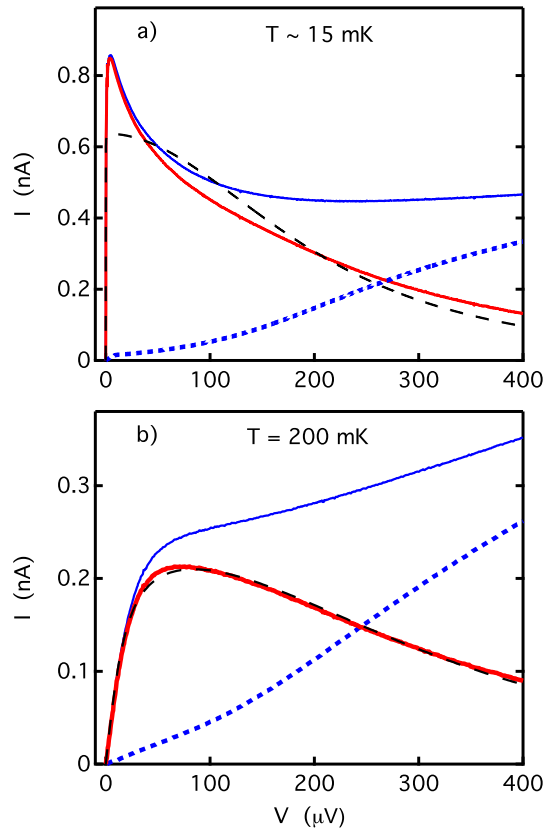}
\end{center}
\caption{(Color online) Subtraction of incoherent tunneling at \nt\ and $d/\ell=1.62$. (a) $T=15$ mK results. Solid blue curve: Four-terminal tunneling $IV$ curve with \bpar=0. Dashed blue curve: $IV$ with the sample tilted to yield \bpar=0.6 T. Red curve: Difference between the solid and dashed blue curves. Dashed black line: Least-squares fit of Eq. (2) to the red curve over the entire voltage range. (b) Same as (a) except at $T=200$ mK.   For the data in this figure, the current source and drain are contacts $2t$ and $1b$, respectively, while the voltage is measured between contacts $3t$ and $4b$.}
\label{HR1}
\end{figure}
In order to compare the HR theory to our experimental results, we restrict our attention to data in which the circuit instabilities discussed in Sec. IV B are not present.  This is essential in order to avoid the data gap at the onset of the resistive portion of the $IV$ curve.  In what follows, we will focus on data acquired at $d/\ell = 1.62$ where the tunneling is weak enough so that no circuit instability occurs and there is no gap in the data.

More significantly, we must also try to remove from the data the contribution of incoherent tunneling processes which are not specific to the \nt\ excitonic phase.  As mentioned in the Introduction, in the absence of interlayer coherence tunneling between 2D electron systems at high magnetic field is characterized by a Coulomb gap around zero bias followed by broad peaks in the tunneling current at voltages comparable to the mean intra-layer Coulomb energy \ec.  While this energy scale is typically much larger than those relevant to the coherent \nt\ phase, these incoherent tunneling features remain visible (especially at high voltages) in the $IV$ curves \cite{spielman00,spielman01} even at low \dl.

Since the incoherent tunneling at \nt\ depends, albeit weakly, on temperature $T$ and effective layer separation \dl, its determination is best accomplished by destroying the coherent portion of the $IV$ curve with these parameters held fixed.  Fortunately, this can be readily done by adding an in-plane magnetic field component \bpar\ to the perpendicular field \bperp\ used to establish \nt.  (This is accomplished by tilting the sample relative to the applied magnetic field.) It is well known that only a relatively small \bpar\ is needed to heavily suppress coherent tunneling at \nt, via a phase-winding mechanism closely related to that which suppresses the critical current in a conventional superconducting tunnel junction \cite{spielman01,tinkham}. 

Figure \ref{HR1} illustrates the subtraction of the \nt\ incoherent tunneling at \dl= 1.62 for $T=15$ and $200$ mK. The solid blue traces are the tunneling $IV$ curves observed at \bpar=0, while the dashed blue traces are the results with \bpar= 0.6 T.  [At this \bpar\ the Josephson-like jump in the tunneling current at $V = 0$ has been reduced to about 1.5\% of its value at \bpar=0, and is nearly invisible in Fig. \ref{HR1}(a). We regard this as sufficient suppression of the coherent part of \nt\ tunneling \cite{bparallel}.]  The red curves are the calculated tunneling $IV$ curves obtained by subtracting the \bpar = 0.6 T data from the \bpar = 0 data.  At both temperatures the contribution of the incoherent tunneling at voltages below about $V \sim 50$ $\mu$V is quite small while by $V \sim 400$ $\mu$V it supplies well more than half of the total tunneling current.

\begin{figure}
\begin{center}
\includegraphics[width=1 \columnwidth] {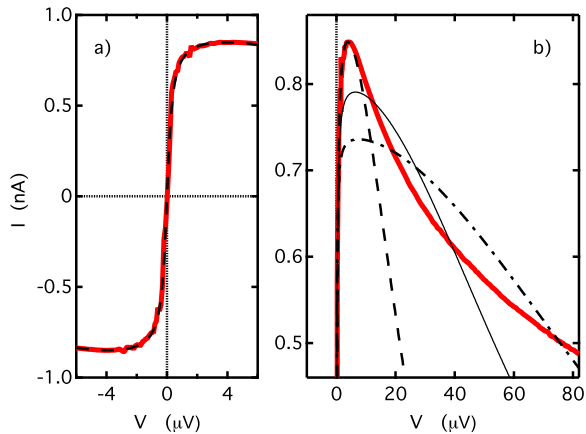}
\end{center}
\caption{(Color online) (a) Red trace: \nt\ tunneling $IV$ curve at $T \sim 15$ mK and \dl= 1.62, with incoherent tunneling contribution subtracted, at very low voltage. Dashed black line: Fit to Eq. (2) over the voltage range $|V| \le 5$ $\mu$V. (b) Red trace: Same $IV$ curve as in (a), displayed over a wider voltage range near the critical current.  The dashed black line is the same fit as in (a). Light solid and dashed-dotted black lines: Fits to Eq. (2) over voltage ranges $|V| \le 50$ and 100 $\mu$V, respectively.}
\label{HR2}
\end{figure}

The black dashed lines in Fig. \ref{HR1} are the results of unweighted least-squares fits of Eq. (2) to the red tunneling $IV$ curves.  The fits are done over the entire voltage range shown in the figure.  For the $T=200$ mK data the fit is quite good, with the extracted fit parameters being $I_0=0.255$ nA, $V_0=394$ $\mu$V, and $\alpha = 0.0401$.  In contrast, the fit to the $T \sim 15$ mK data is poor, especially where the ``supercurrent'' branch meets the resistive portion of the $IV$ curve at the critical current.  In this case the fit parameters are $I_0=0.638$ nA, $V_0=252$ $\mu$V, and $\alpha = (7 \pm 4) \times 10^{-5}$.  This very small and highly uncertain value of $\alpha$ reflects the extreme steepness of the ``supercurrent'' branch at $T \sim 15$ mK.

Better fits to the ``supercurrent'' branch of the $IV$ curve can be obtained by reducing the voltage domain of the fit.  This is demonstrated in Fig. \ref{HR2}a where the near-perfect fit of Eq. (2) to the $T = 15$ mK tunneling $IV$ curve over the narrow voltage range $|V| \le 5$ $\mu$V is shown.   Figure \ref{HR2}(b) shows however, that this same fit fails at voltages in excess of about 10 $\mu$V.  Expanding the fit domain [to $|V| \le 50$ and 100 $\mu$V, as shown in Fig. \ref{HR2}(b)] improves the fit at higher voltages, but only at the expense of a poorer fit near the cusp in the $IV$ curve at the critical current. 

Apparently, the transition from the ``supercurrent'' branch to the resistive branch at very low temperature is not captured by the HR theory \cite{hyart11}.  We emphasize that this conclusion does not rely on the efficacy of the incoherent tunneling subtraction since the latter contribution to the tunneling current is negligible at these low voltages.  At higher temperatures, the HR theory does a much better job.  This situation was actually anticipated by HR who point out that at very low temperatures, where $\alpha$ is small, their perturbative approach to tunneling may no longer be justified \cite{hyart11}.

\section{The ``supercurrent'' branch}
\label{scbranch}
Perhaps the most dramatic feature of the tunneling $IV$ curves at \nt\ is the nearly vertical ``supercurrent'' branch observed around zero bias at very low temperatures.  This feature, which strongly resembles the dc Josephson effect, has been the subject of intense interest for over a decade.  In this section we describe the results of measurements in this portion of the tunneling $IV$ curve, restricting our attention to the vicinity of zero tunneling current.

\subsection{Experimental issues}
For the tunneling $IV$ data presented above, the four-terminal voltage difference is measured between top and bottom layer contacts which are physically separated along the edge of the Corbino annulus.  Consequently, this voltage is ``diagonal'' and will, in general, contain both inter- and intra-layer contributions.  Fortunately, the intralayer part of this voltage can be separately measured and then subtracted from the diagonal voltage to yield the purely interlayer voltage at either contact. For example, the inter-layer voltage at contact 2 can be deduced from measurements of the diagonal voltage $V_{2t,3b}$ and the intra-layer voltage $V_{2b,3b}$, via Kirchoff's law: $V_{2t,2b}=V_{2t,3b}-V_{2b,3b}$.  (Note that Kirchoff's law implies that the same inter-layer voltage can be determined in two independent ways: $V_{2t,2b}=V_{2t,3b}-V_{2b,3b}=V_{2t,3t}-V_{3t,2b}$.)  This procedure for subtracting the intra-layer voltage is only necessary on the ``supercurrent'' branch, where the inter-voltage is extremely small.  In the resistive portion of the tunneling $IV$, the intra-layer voltages are negligible in comparison to the inter-layer voltages.

When the coherent \nt\ state is well-developed, the inter-layer voltages along the ``supercurrent'' branch are sufficiently small (typically sub-$\mu$V) that dc amplifier drift makes ac lock-in detection essential.  In what follows, we describe inter-layer voltage measurements performed using purely ac $current$ excitation, with $I_{ex} = 0.2$ nA at 13 Hz.  To avoid spurious signals arising from capacitive effects, we found it necessary to inject the current using a ratio-transformer bridge in order to reduce the common-mode voltage of the tunnel junction.  In addition to recording the inter- and intra-layer voltages which developed, the current was also measured so that small deviations from perfect current bias could be accounted for.

\begin{figure}
\begin{center}
\includegraphics[width=1 \columnwidth] {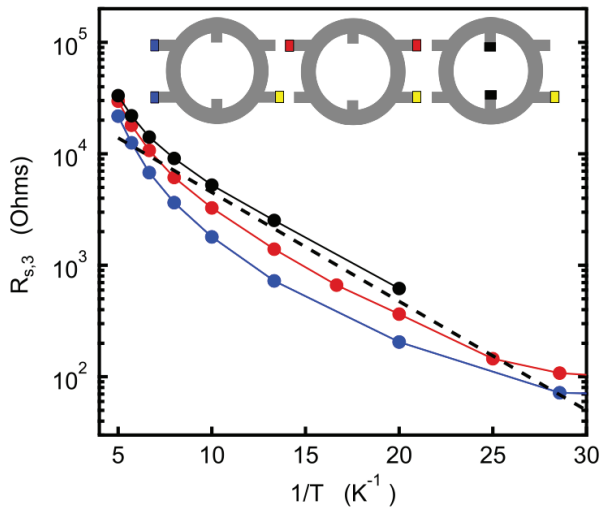}
\end{center}
\caption{(Color online) Resistance $R_s$ of the ``supercurrent'' branch vs. inverse temperature at \nt\ and $d/\ell = 1.49$.  $R_s$ is determined from the interlayer voltage at contact 3 (yellow square in the device schematics) arising in response to ac current bias for three different source/drain configurations (blue, red, and black squares in the device schematics).  For comparison, the dashed straight line indicates simple thermal activation.}
\label{rtun}
\end{figure}

\subsection{Resistance of the ``supercurrent'' branch}
The ratio of the ac inter-layer voltage at contact $j$ to the ac tunneling current provides a measure of the resistance $R_{s,j}$ of the ``supercurrent'' branch around zero bias.  Figure \ref{rtun} displays the observed temperature dependence of $R_{s,3}$ at \nt\ and $d/\ell = 1.49$.  (Very similar results are obtained for the other contacts in the device.) Data for three different current source/drain configurations are shown: $I_{2t,1b}$, $I_{4t,1b}$, and $I_{5t,6b}$.   In all three configurations, $R_{s,3}$ drops rapidly as the temperature is reduced.  For comparison, the dashed line shows an assumed thermally activated dependence, $R_s \sim \exp(-\Delta/2T)$, with $\Delta = 450$ mK.  This value of $\Delta$ is essentially the same as the \nt\ energy gap we deduce from conventional in-plane transport measurements at this \dl.

The data in Fig. \ref{rtun} demonstrate that the resistance $R_s$ of the ``supercurrent'' branch remains finite down to at least $T = 30$ mK. (This appears to remain true at still lower temperatures, but measurement difficulties suggest further work is needed in this regime.)  Moreover, the resistance depends significantly on the location of the current source and drain contacts.  This stands in sharp contrast to the critical tunneling current $I_c$ which, as Fig. \ref{contact_indep} proves, is independent of the source/drain contact configuration.  

Our data suggest that by $T = 30$ mK, the resistance of the ``supercurrent'' branch falls to around $R_s \sim 100$ $\Omega$.  By this measure, the zero bias tunneling conductance $G(0) = R^{-1}_s \sim 250~e^2/h$ at \nt\ is some 6000 times larger than it is at zero magnetic field where tunneling is essentially a single-particle phenomenon \cite{zeroBtun}.  This comparison makes plain the highly collective nature of tunneling at \nt.

\section{Discussion of Results}

The similarity between the tunneling $IV$ characteristics in coherent \nt\ bilayers and superconducting Josephson junctions was first noted by Spielman {\it et al.} via two-terminal measurements on weakly tunneling samples of small area \cite{spielman00,spielman01}.  Subsequent measurements by Tiemann {\it et al.} \cite{tiemann08a,tiemann09} displayed the striking difference between two- and four-terminal tunneling experiments on large area strongly tunneling samples.  This difference arises both from extrinsic series resistance in the measurement circuit and from an intrinsic quantum Hall contact resistance \cite{rossi05,pesin11}.  Aside from reproducing Tiemann's findings, we have shown that the instabilities and hysteresis in the $IV$ are not deeply related to the many-body physics of the coherent \nt\ bilayer but instead arise from relatively mundane circuit effects which become important when tunneling is strong \cite{hyart11}.

In spite of the instabilities and hysteresis effects observed in strongly tunneling situations, we have shown that the maximum, or critical, tunneling current $I_c$ is unaffected by them and is a genuine feature of tunneling in coherent \nt\ systems.  Our results suggest that $I_c$ is an intrinsic property of the system, being independent of the contact configuration used to make the tunneling measurement and limiting the total tunneling current even when two source-drain contact pairs on opposite rims of the Corbino annulus are used to inject current.  These findings support the prior evidence that tunneling in coherent \nt\ bilayers is a bulk phenomenon, with the maximum tunneling current proportional to the area of the 2D system.

This scaling with area is surprising since one expects, in an ideal, disorder-free \nt\ system, tunneling currents to be confined to within a short distance of the source and drain contacts.  This distance, $\lambda_J = 2 \ell \sqrt{\pi \rho_s/\Delta_{SAS}}$ (with $\rho_s$ the pseudospin stiffness in the coherent phase, and $\Delta_{SAS}$ the single-particle tunnel splitting in the double quantum well), is analogous to the Josephson penetration length in superconducting junctions \cite{tinkham} and is estimated to be in the $\mu$m range, far smaller than the mm-scale dimensions of our Corbino ring.  Presumably, as has been suggested \cite{fertig05,eastham10}, disorder in realistic \nt\ samples leads to an effective coarse-grained average of $\lambda_J$ which is much larger than the ideal value.  

Consistent with prior results \cite{eisenstein03,tiemann09}, we find that $I_c$ depends strongly on the effective layer separation \dl\ at \nt, rising from zero at $d/\ell \approx 1.8$ to $I_c \gtrsim 2$ nA at $d/\ell = 1.39$ and $T \sim 15$ mK.   Surprisingly, however, we find that the temperature dependence of the normalized critical current $I_c/I_{c,max}$ (with $I_{c,max}$ the $T \rightarrow 0$ value of the critical current) is nearly independent of \dl.  To the extent that this is precisely the case, 
our data suggest that the critical current is of the form $I_c(T,d/\ell) = g(d/\ell) \times f(T/T_0)$, with $g$ a function of \dl\ alone and $f$ a function of $T/T_0$ alone, with the characteristic temperature $T_0$ independent of \dl.  It seems peculiar that the temperature-induced suppression of coherent interlayer tunneling would be independent of \dl, since this parameter reflects the relative strength of inter- and intra-layer Coulomb interactions and governs the onset of all the exotic transport phenomena observed at \nt.  One speculative possibility is that $T_0$ is determined entirely by disorder, the bare unscreened potential of which is fixed, independent of the electron density in the quantum wells.  We emphasize however, that $I_c/I_{c,max}$ is not precisely independent of \dl, with deviations visible at the largest effective layer separations.  Moreover, at higher temperatures, where the maximum tunneling current occurs at a significant non-zero voltage, incoherent processes contribute substantially to the tunneling current.  Subtracting these contributions, in a manner similar to that described in Sec. \ref{HR}, will likely enhance the \dl\ dependence of $I_c/I_{c,max}$.

At elevated temperatures we find that the recent phenomenological theory \cite{hyart11} of the tunneling lineshape agrees reasonably well with our observations, provided that the incoherent contribution to the tunneling current is subtracted.  At lower temperatures this does not seem to be the case.  The experimentally observed tunneling $IV$ curve then shows a very pronounced cusp-like feature where the ``supercurrent'' branch meets the resistive portion of the $IV$ characteristic.  Fits of the Hyart-Rosenow \cite{hyart11} theory to the data do not capture this cusp.

While the maximum {\it two-terminal} tunneling conductance at \nt\ is expected to remain finite (and of order $e^2/2h$) down to the lowest \dl\ and temperature \cite{rossi05,pesin11}, the situation is less clear regarding the four-terminal conductance around zero bias.  Our results, summarized in Fig. \ref{rtun}, clearly show that the resistance of the ``supercurrent'' branch falls rapidly with temperature, but remains finite certainly down to $T \sim 30$ mK and likely to even lower temperature.  Experimentally, therefore, it does not appear that the condensate phase $\phi$ is time-independent along the ``supercurrent'' branch, even if its precession rate is vastly less than it is on the resistive portion of the $IV$ curve.  In our view, this is not surprising.  The data presented here and elsewhere have shown that tunneling at \nt\ occurs throughout the bulk of the 2DES.  In order for this to be true, in-plane transport must be involved.  In spite of the very small conductivity of the bulk for net charge transport (the coherent \nt\ bilayer is a conventional quantum Hall system in this regard), it is virtually transparent to neutral exciton transport within the condensate \cite{finck11,nandi12}.  Since the latter consists of counterflowing charge currents in the two layers, it can readily relax the anti-symmetric charge defects created by tunneling.  As a corollary therefore, any dissipation accompanying exciton transport can be expected to also show up in the tunneling $IV$ curve.  At the same time, the local interlayer voltage created by such dissipation would naturally depend on the spatial distribution of the exciton transport.  This could account for the dependence of the resistance $R_s$ of the ``supercurrent'' branch on the relative locations of the source, drain, and interlayer voltage contacts shown in Fig. \ref{rtun}.

Dissipation in counterflow transport has, to date, been most sensitively studied in experiments employing Hall bar geometries \cite{kellogg04,tutuc04,wiersma04}.  All of these experiments suggest that dissipation remains finite down to the lowest temperatures.  Moreover, the experiments show that for small currents, the observed response is linear; there is as yet no clear evidence for the predicted non-linear effects due to current-induced vortex pair ionization \cite{moon95}.  

Interestingly, the data shown in Fig. \ref{rtun} suggest that the temperature dependence of the resistance of the ``supercurrent'' branch is not terribly different from simple thermal activation and, as mentioned above, is in fact comparable to that observed for conventional in-plane charge transport.  (The data shown in Fig. \ref{rtun} do show more curvature than is typically found in the temperature dependence of conventional charge transport.) Similarity between the observed dissipation in counterflow and ordinary quantum Hall transport at \nt\ has been noticed before, in both electron-electron \cite{kellogg04} and hole-hole \cite{tutuc04} bilayers.  This similarity suggests that dissipation in counterflow at low temperatures is governed by the ordinary quantum Hall energy gap for charged quasiparticles.  If, as has been suggested \cite{fertig03,fertig05,eastham09,jpe04,huse05}, the motion of single, disorder-induced, unpaired vortices is responsible for the observed dissipation in neutral exciton transport, these observations suggest that the same process may also be responsible for dissipation in ordinary charge transport.  This contrasts with the usual theoretical assumption that the dominant charged excitation at \nt\ is a meron-antimeron pair.  While carrying electrical charge ($\pm e$), such objects are vortex neutral and thereby unable to unwind an excitonic superflow. 

\section{Conclusion}
An extensive set of tunneling measurements on strongly correlated bilayer 2D electron systems at \nt\ has been reported here.  Our results are consistent with the many prior experimental observations but also provide several different perspectives.  Prominent among these are the observations of an unexpected scaling behavior the maximum, or critical, tunneling current and a precise independence of the critical current on contact configuration.  Our findings also provide a detailed view of the temperature dependence of the tunneling resistance along the ``supercurrent'' branch of the $IV$ curve and allow for a sensitive comparison to recent theories of the tunneling lineshape.

\begin{acknowledgements}
It is a pleasure to thank Ron Lifshitz, Allan MacDonald, Dmytro Pesin, Johannes Pollanen, Bernd Rosenow, Ian Spielman, and Jung-Jung Su for numerous helpful interactions.  The Caltech portion of this work was funded in part by the Institute for Quantum Information and Matter, an NSF Physics Frontiers Center with support of the Gordon and Betty Moore Foundation through Grant No. GBMF1250, and by NSF Grant No. DMR-1003080.  The work at Princeton was partially funded by the Gordon and Betty Moore Foundation through Grant No. GBMF2719, and by the National Science Foundation MRSEC-DMR-0819860 at the Princeton Center for Complex Materials.
\end{acknowledgements}


\begin{references}
\bibitem{rext} The slight decline of the maximum tunneling current with $R_{ext}$ shown in Fig. \ref{instab}(c) is not understood at present.  We note however that the area of the hysteresis loops evident in Fig. \ref{instab}a grows rapidly with increasing $R_{ext}$ and that the higher current state is at most metastable.
\bibitem{zeroBtun}  Direct measurements of the low temperature tunneling conductance at $B=0$ and the same balanced layer densities as in Fig. \ref{rtun}, show a clean symmetric peak at zero bias with height $G(0) = 1.5 \times 10^{-6}$ $\Omega^{-1}$.  The width of the tunneling peak, which provides a measure of the electronic quantum lifetime, is about 90 $\mu$V.
\bibitem{equation} We have added the factor of $\pi$ in the denominator of Eq. 1 to the formula given by Hyart and Rosenow (Ref. 60).  By so doing, the maximum tunneling current, in the limit $\alpha \rightarrow 0$, becomes $I_0$ instead of $\pi I_0$.
\bibitem{bparallel} While a higher \bpar\ reduces the critical current still further, additional effects come into play.  For example, the incoherent tunneling itself is eventually suppressed by the in-plane field and we wished to minimize this effect (Ref. 70) (it is less than 10\% for the data presented here and has in any case been taken into account in Figs. \ref{HR1} and \ref{HR2}). In addition, we have discovered an intriguing enhancement of the incoherent tunneling Coulomb gap which occurs at large \bpar.  While irrelevant here, we defer discussion of this effect to a subsequent publication.
\end{references}
\end{document}